# Spatially multiplexed picosecond pulse-train generation in a 6-LP-mode fiber based on multiple four-wave-mixings

**H. ZHANG,[1] M. BIGOT-ASTRUC,[2] P. SILLARD,[2] AND J. FATOME[1],***

[1] *Laboratoire Interdisciplinaire Carnot de Bourgogne, UMR 6303 CNRS-Université Bourgogne Franche-Comté, 9 av. A. Savary, 21078 Dijon Cedex, France*
[2] *Prysmian Group, Parc des Industries Artois Flandres, Haisnes 62092, France*
**Julien.Fatome@u-bourgogne.fr*

**Abstract:** We report on the generation of four spatially multiplexed picosecond 40-GHz pulse trains in a km-long 6-LP multimode optical fiber. The principle of operation is based on the parallel nonlinear compression of initial beat-signals into well separated pulse trains owing to intra-modal multiple four-wave mixings. A series of four 40-GHz dual-frequency beatings at different wavelengths are simultaneously injected into the $LP_{01}$, $LP_{11}$, $LP_{02}$ and $LP_{12}$ modes of a 1.8-km long graded-index few-mode fiber. The combined effects of Kerr nonlinearity and anomalous chromatic dispersion lead to the simultaneous generation of four spatially multiplexed frequency combs which correspond in the temporal domain to the compression of these beat-signals into picosecond pulses. The temporal profiles of the output pulse-trains demultiplexed from each spatial mode show that well-separated picosecond pulses with negligible pedestals are then generated.

## 1. Introduction

The ability to generate picosecond pulse trains at high repetition rates is a fundamental issue which can find numerous applications in optical communications, optical sampling, component characterization, clock generation, metrology or spectroscopy. Despite the giant progress of opto-electronic components in terms of bandwidth, it remains challenging to directly generate picosecond pulses at high repetition rates. To overcome the limit of electronic bandwidth, all-optical techniques have been proposed. Active mode-locked lasers is an attractive solution to generate picosecond pulses but it remains a costly and sensitive device [1-5]. In parallel, several all-optical cavity-free techniques have been proposed [6-21]. One of the main principles is based on the linear or nonlinear reshaping of an initial sinusoidal modulation into well-separated pulses. The first linear family of solutions is based on the direct temporal phase modulation that is then converted into an intensity modulation through the insertion of a spectral quadratic phase owing to a dispersive element. Thanks to this technique, picosecond pulses at repetition rates of several tens of GHz have been successfully generated [6-10]. The second technique is based on the nonlinear compression of the initial temporal beating into a train of well separated pulses thanks to the combined effect of Kerr nonlinearity and chromatic dispersion within one or an arrangement of optical fibers. To this aim, adiabatic soliton compression, dispersion decreasing fiber, comb-like and step-like dispersion profiles have been investigated [11-16]. However, these techniques often require a relatively complicated experimental setup and a careful management of the chromatic dispersion all along the fiber length.
A specific configuration of this nonlinear compression technique is based on multiple four-wave mixing (MFWM). This process occurs in a single anomalous dispersive fiber and has been proved to be an attractive and efficient technique to generate very high-repetition-rate pulse trains, combining both stability and simplicity [17-21]. In this new contribution, we demonstrate the simultaneous spatially multiplexed generation of four 40-GHz picosecond pulse trains in a few mode fiber based on MFWM. To this aim, we exploit a 6-LP few mode fiber in which four different group of modes ($LP_{01}$, $LP_{11}$, $LP_{02}$

& $LP_{21}$ and $LP_{12}$ & $LP_{31}$) act as four individual pulse generators. More precisely, by simultaneously injecting four beat-signals with unmatched group velocity central wavelengths into four different spatial modes, the parallel pulse-train generations in all the modes can be efficiently performed without significant degradations. The performance of pulse compressions in each spatial mode has been experimentally investigated. Moreover, the cross-talks induced by the intermodal cross-phase modulation between the fundamental mode ($LP_{01}$) and the higher-order modes have been examined. Numerical simulations based on two-mode nonlinear Schrödinger equation (NLS) are in good agreement with our experimental results.

## 2. Principle of operation

The principle of operation is schematically described in Fig. 1(a), where four dual-frequency beat-signals (black arrows) are injected into four different spatial modes of the 6-LP-mode fiber. The key point of this experiment is that the central wavelength of each beat-signal has been carefully selected in order to vanish any inter-modal nonlinear interactions, so as to avoid the nonlinear cross-talk between the different channels through cross-phase modulation. More precisely, as can be seen on the measured dispersion curves of relative inverse group velocity (RIGV) reported in Fig. 1(b), the central wavelengths for the $LP_{11}$, $LP_{02}$ and $LP_{12}$ modes with respect to the $LP_{01}$ have been chosen to be not group-velocity matched, thus avoiding nonlinear coupling and cross-talk between the modes through intermodal nonlinear interactions [22-25]. Indeed, the same group velocities of the four modes are obtained for a wavelength offset of $\Delta\lambda \approx 16$ nm, 25 nm, and 31 nm, respectively. Consequently, a close to 4-nm offset has been here applied in between the different modes under study. Subsequently, each mode acts as an individual nonlinear channel that develops its own intra-MFWM due to the interplay between the Kerr effect and chromatic dispersion, leading to a broad concatenated output frequency comb. Meanwhile, in the temporal domain, this parallel processing leads to the nonlinear compression of the four initial beat-signals into well-separated picosecond pulse trains.

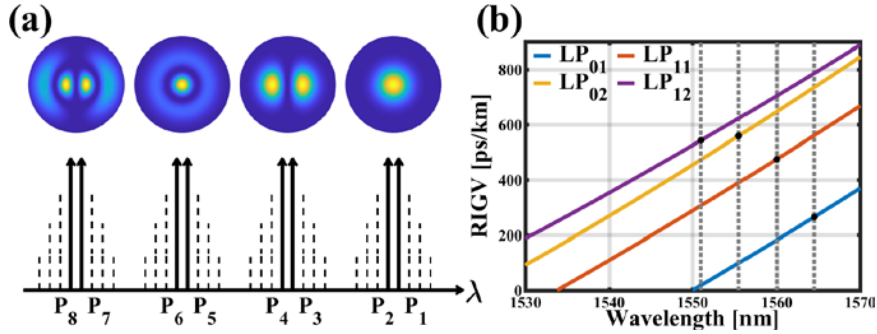

**Fig. 1.** (a) The excited modes at the input of the 6-LP-mode fiber, from right to left: $LP_{01}$, $LP_{11}$, $LP_{02}$ and $LP_{12}$. (b) Experimental measurements of the relative inverse group velocity (RIGV) vs wavelength for each mode under study. The selected central wavelengths in each mode are denoted by grey dashed lines.

## 3. Experimental setup

In order to experimentally demonstrate the simultaneous nonlinear compression of four spatially multiplexed 40-GHz sinusoidal beatings in our few-mode fiber, we have implemented the experimental setup depicted in Fig. 2. Four tunable external cavity lasers (ECL) in the C-band are used for this experiment. The four different continuous-waves (CW) are first combined and phase-modulated around 100 MHz in order to prevent any Brillouin backscattering in the fiber undertest. A LiNbO$_3$ intensity modulator (IM1), driven around its zero-transmission point by a 20-GHz external RF clock is then inserted to generate the initial 40-GHz beatings. In order to increase the peak power in the fiber under-test, while keeping the average power to the Watt level, the four beatings are then sliced into the temporal domain owing to a second intensity modulator (IM2) driven by electrical square pulses of 500 ps and a duty-cycle of 1:4. The resulting signal is then boosted by a first Erbium-doped fiber amplifier (EDFA) at an average power of 20 dBm. Furthermore, a programmable liquid-crystal-based optical filter (waveshaper) is used to apply a frequency grid centered on each individual dual pumps to remove undesirable residual spectral bands located at 20 GHz. The four initial beatings are then frequency demultiplexed by means of an arrayed waveguide grating (wavelength demultiplexer) and amplified thanks to four single-mode EDFAs. The input initial beatings are then coupled into the FMF fiber by means of a specifically designed 10-mode spatial multiplexer provided by Cailabs based on a multi-plane light conversion technology [26]. Polarization controllers are also inserted in each pump optical path in such a way to maximize the excitation efficiency for each spatial mode. The fiber undertest consists in a 1.8-km long 6-LP-mode graded-index fiber (10 spatial modes including degenerate modes) manufactured by Prysmian group with a core diameter of 22.5 µm [25, 27]. The specific fiber length is selected based on the empirical relation $L_{opt} = \frac{4\pi^2}{14|\beta_2|\Omega^2}$ firstly found for the optimal compression of dual-frequency beat-waves in a single-mode fiber (here for the fundamental mode), where $\beta_2$ represents the fiber dispersion and $\Omega$ is the beat-frequency in units of rad/s [18]. The effective areas ($A_{eff}$) of each mode under-study around 1550 nm are: 75 µm$^2$ for the fundamental mode LP$_{01}$, 100 µm$^2$ for LP$_{11a}$ and LP$_{11b}$, 160 µm$^2$ for LP$_{02}$ and 170 µm$^2$ for LP$_{12a}$ and LP$_{12b}$. The dispersion curves of each mode have been measured by the time of flight method [25], which lead to a chromatic dispersion of 17.5 ps/nm/km for the modes LP$_{12}$, 18 ps/nm/km for LP$_{01}$ and LP$_{11}$, and 19 ps/nm/km for LP$_{02}$, respectively. The losses for all the modes are below 0.25 dB/km, and the maximum differential mode group delay (DMGD) between the modes is lower than 550 ps/km. Moreover, for temporal characterizations, the injection and demultiplexing conditions (polarization controller and fiber stress before demultiplexing operation) were optimized for each mode undertest. The total losses of the system were measured at an average value of 9.3 dB for all the modes while the average cross-talks between the different groups of modes were found better than 20 dB. Note finally, that at the output of the system, a tunable filter is also used to remove the leakage from other spatial modes.

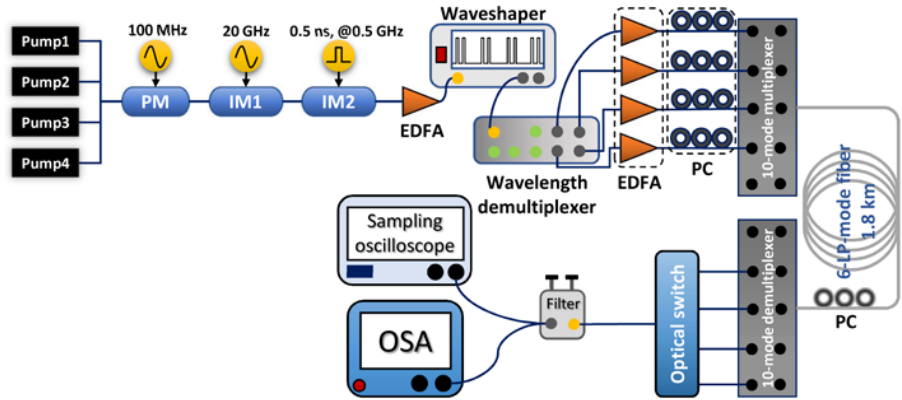

**Fig. 2.** Experimental setup. PM phase modulator, IM intensity modulator, PC polarization controller, EDFA Erbium-doped fiber amplifier, OSA optical spectrum analyzer.

## 4. Experimental results

As a preliminary test, a 40-GHz beating with the central wavelength of 1564.5 nm is injected into the fundamental mode of the 6-LP-mode fiber, while the output spectrum and temporal profile have been measured as a function of the injected average power (given before the spatial multiplexer input). The results are shown in Figs. 3(a) and 3(b). It can be clearly seen from Fig. 3(a) that multiple four-wave-mixing sidebands are generated with increasing power, and the corresponding temporal profiles shown in Fig. 3(b) are gradually compressed. As can be observed in Figs. 3(c) and 3(d), similar nonlinear dynamics can be obtained by injecting the beating into the first higher-order mode $LP_{11a}$. However, since the effective area of this mode is larger than the fundamental mode, the Kerr nonlinearity is smaller and the required power thus increases. The compressions of input beat-waves are also observed in other higher-order modes $LP_{02}$ and $LP_{12a}$, and the efficiency of the compression process for each mode with respect to its normalized nonlinear fiber length (defined here as $L/L_{nl} = \gamma P L$ with $P$ the input beat-wave power in each mode) is plotted in Fig. 3(e). Here the nonlinear Kerr coefficient is obtained by $\gamma = n_2 \omega / (c A_{eff})$ with $n_2 = 2.6 \times 10^{-20}$ m$^2$/W the silica Kerr nonlinear refractive index and $c$ the light velocity in vacuum. Because of the strong linear coupling between degenerate modes along the km-scale fiber, each higher order mode is described by the superposition of all the degenerate modes within the same spatial group, i.e. $E_{11} = E_{11a} \pm i E_{11b}$ for $LP_{11}$ with $E$ representing the transverse electric field of the LP modes [23]. Then the effective areas for $LP_{11}$, $LP_{02}$, and $LP_{12}$ are given as $A_{eff}^{11} = 160$ μm$^2$, $A_{eff}^{02} = 231$ μm$^2$ and $A_{eff}^{12} = 318$ μm$^2$. We can observe that similar performance with a minimum temporal duration of 4.9 ps, corresponding to a duty-cycle of 1:5, is achieved in the $LP_{01}$ and $LP_{11}$ modes, while the compressions in the $LP_{02}$ and $LP_{12}$ modes are largely reduced due to mode coupling. In Fig. 3(f), we have compared the output temporal profiles of each mode at the optimal compression point. We note that for each mode, the initial beating is well compressed under the combined action of Kerr nonlinearity and chromatic dispersion, though the compression becomes less efficient for the higher order modes $LP_{02}$ (7.7 ps) and $LP_{12}$ (9.2 ps), probably due to their larger effective areas as well as the random coupling between degenerate modes which can deteriorate the intra-MFWM processes.

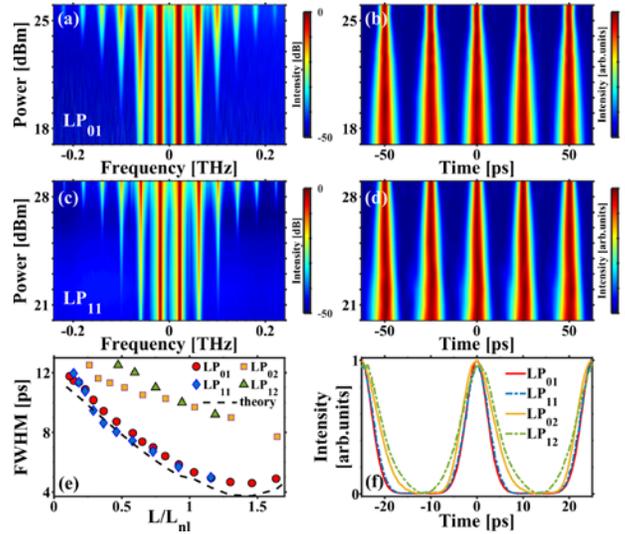

**FIG. 3.** Spectral and temporal evolution of the output pulse-train monitored in the fundamental mode $LP_{01}$ (a, b) and the first higher-order mode $LP_{11}$ (c, d) as a function of the injected power. (e) Output pulse duration (full-width-half-maximum) as a function of the normalized nonlinear fiber length $L/L_{nl}$ (see text) for the $LP_{01}$, $LP_{11}$, $LP_{02}$ and $LP_{12}$ modes together with the simulation results (the four modes involved in our experiment provide the same dashed-lines). (f) Intensity profile recorded at the optimal compression point for $LP_{01}$ (red solid line), $LP_{11}$ (blue dot-dashed line), $LP_{02}$ (yellow solid line) and $LP_{12}$ (green dot-dashed line) are shown in.

Next we investigate the concurrent compressions of two beat-waves injected into two different spatial modes. To this aim, two 40-GHz beatings are respectively injected into the $LP_{01}$ and $LP_{11a}$ modes, with each power set at its optimal compression point. The spectral and temporal profiles demultiplexed and measured at the output of the 6-LP mode fiber are shown in Fig. 4(a-d) as red dotted lines for the $LP_{01}$ (a-b) and $LP_{11a}$ (c-d) modes, with the wavelength offset between the two beat-waves corresponding to the group-velocity matching condition ($\Delta\lambda \approx 16$ nm) (same inverse group velocity) [25]. For comparison the frequency comb as well as the optimally compressed pulse train, obtained with each beating separately switched on, are also plotted with black solid lines. It can be easily noticed that strong perturbations, both for the spectral and temporal profiles, appear when both waves are group-velocity matched. In this case, although the frequency combs generated in both modes are further broadened by the intermodal nonlinear processes, their temporal profiles become severely distorted especially for $LP_{11a}$ with the emergence of large pedestals.

To get a better understanding of the multiple four-wave-mixing process and the impact of intermodal nonlinear interactions in the few mode fiber, especially cross-phase modulation, we use the two-mode coupled NLS to simulate the simultaneous compressions of two beat-waves. The simplified scalar NLS neglecting the higher-order (third-order and higher) dispersion terms and also the Raman effect is employed as [28]:

$$\frac{\partial E_p}{\partial z} = \left[\frac{\delta_{pq}}{2}\frac{\partial}{\partial t} - \frac{i}{2}\beta_{2p}\frac{\partial^2}{\partial t^2} + i\frac{n_2\omega}{c}\left(f_{pp}|E_p|^2 + 2f_{pq}|E_q|^2\right)\right]E_p, \quad (1)$$

$$\frac{\partial E_q}{\partial z} = \left[-\frac{\delta_{pq}}{2}\frac{\partial}{\partial t} - \frac{i}{2}\beta_{2q}\frac{\partial^2}{\partial t^2} + i\frac{n_2\omega}{c}\left(f_{qq}|E_q|^2 + 2f_{pq}|E_p|^2\right)\right]E_q, \quad (2)$$

where the reference frequency $\omega$ is set at the average wavelength of the two modes, $\delta_{pq} = \beta_{1p} - \beta_{1q}$ is the group velocity mismatch with $\beta_{1i}$ the inverse of the group velocity in the mode ($i=p, q$), $\beta_{2i}$ is the second-order dispersion coefficient. $1/f_{pp}$ and $1/f_{qq}$ are the effective areas of the two interacting modes with $1/f_{pq}$ the effective coupling area between them. The input beat signals for each mode can be expressed as:

$$E_p(z=0,t) = \sqrt{P_p/2}\exp(-i\Delta\omega_p t)\times\left[\exp(-i\Omega t/2) + \exp(+i\Omega t/2)\right], \quad (3)$$

$$E_q(z=0,t) = \sqrt{P_q/2}\exp(-i\Delta\omega_q t)\times\left[\exp(-i\Omega t/2) + \exp(+i\Omega t/2)\right], \quad (4)$$

with $\Delta\omega_i = \omega_i - \omega$ the frequency-detuning from the reference frequency and $P_i$ the injected power in each mode.

The temporal evolution of the input beat-wave has been first simulated as a function of the injected power in each mode separately, and the retrieved pulse duration (FWHM) as a function of the normalized nonlinear fiber length was plotted for all the modes with black dashed-lines in Fig. 3(e). Note that due to the weak difference of chromatic dispersion between the spatial modes involved in our experiment, all the curves superimpose. It is clearly seen that the simulated compression vs nonlinear length curve is in good consistence with the experimental measurements for the fundamental mode and the $LP_{11}$ mode, while large deviations from the simulation results are observed for the measurements on the $LP_{02}$ and $LP_{12}$ modes. We attribute this discrepancy to the random linear coupling between the degenerated modes in higher order groups of modes which is neglected into our simplified model. Next the nonlinear cross-talk between the concurrent compressions in the $LP_{01}$ and $LP_{11}$ modes is simulated, by injecting simultaneously both beat-waves at corresponding optimal powers in each mode. The central wavelength of the beat-wave in $LP_{11}$ is scanned from 1545 nm to 1562 nm while the central wavelength in $LP_{01}$ is fixed at 1564.5 nm. The temporal profiles of the compressed beat-waves are monitored as a function of the wavelength offset and depicted in Fig. 4(e) for $LP_{01}$ and 4(f) for $LP_{11}$. It is clearly demonstrated that the compressions in both modes are essentially independent with each other, except for a specific and narrow wavelength offset range around the group-velocity matching conditions ($\Delta\lambda \approx 16$ nm) between the two modes for which intermodal cross-phase modulation clearly provides severe distortions on the temporal profile.

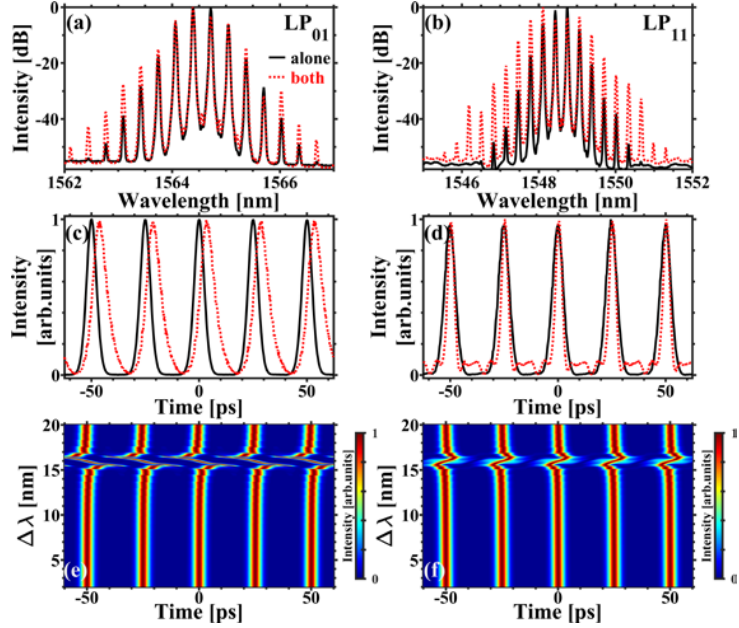

**Fig. 4.** The output spectral and temporal profiles measured at the optimal compression point for $LP_{01}$ (a)&(c) and $LP_{11a}$ (b)&(d), with the black solid lines representing the case of separate injections of each beat-wave and the red dotted lines denoting the output in the simultaneous compression of both beat-waves in the group velocity matching condition ($\Delta\lambda \approx 16$ nm). The output temporal profiles at the optimal compression points for each mode are simulated and depicted in (e) and (f), as a function of the wavelength offset $\Delta\lambda$ between the two modes (see text).

Based on the above experimental and simulation results, we can now conclude that the key factor determining the nonlinear cross-talks between different modes and nonlinear channels is the central working wavelength of each channel. More precisely, negligible cross-talk and nonlinear impairments can be achieved by setting the injected wavelengths in each mode far from the group-velocity matched wavelengths between all the active modes. To further demonstrate experimentally that point, we have spatially multiplexed the four 40-GHz beat-signals into the FMF under-test. The wavelength and injected power (given at the spatial multiplexer input) for each mode are respectively, [1562.5 nm, 25 dBm for $LP_{01}$], [1558.5, 27 dBm for $LP_{11a}$], [1554.5 nm, 33 dBm for $LP_{02}$] and [1550.5 nm, 33 dBm for $LP_{12a}$]. Note that similar results can be achieved in other degenerate modes.

Fig. 5a depicts the total spectrum recorded at the output of the FMF obtained by summing the four demultiplexed spatial channels. We can clearly observe the simultaneous generation of spatially multiplexed Kerr frequency combs at a repetition rate of 40 GHz, thus demonstrating the efficiency of the intra-modal MFWM in each excited spatial mode. Figures 5(b-e) illustrate the corresponding intensity profiles recorded in each demultiplexed spatial mode. The temporal profile has been here measured thanks to an optical sampling oscilloscope with a 1 ps resolution (PSO-101). Note that the presence of a 'zero' level is due to the 1:4 duty-cycle imposed to the initial beatings. We can observe that for each spatial mode, well-separated pulses are generated with a symmetric shape and low residual background or pedestal. Nevertheless, we can notice that the compression efficiency in higher order modes decreases due to a larger effective area and thus a weaker nonlinearity. Note also that the input state-of-polarization (SOP) for each spatial mode has been carefully adjusted, especially for highly degenerated modes, in order to optimize the temporal profile. Indeed, due to random mode coupling within the fiber under test as well as polarization dependent performance of our spatial multiplexer and demultiplexer, we adjusted the input SOPs of each spatial channel in such a way to optimize the energy injected into a specific degenerated mode, thus maximizing the efficiency of the intra-modal MFWM process. Nevertheless, fluctuations over time of the quality of the generated pulse train have been observed, particularly in the highest-order degenerate modes ($LP_{12}$). This behavior is attributed to the evolution of the linear mode coupling which randomly spreads the energy in the whole group of modes due to external perturbations (temperature, vibration…). However, these fluctuations occur in a time scale of several minutes and can be compensated by readjusting the polarization controllers. We have also compared in Fig. 5 the possible impact of the cross-talks between the different modes under study. More precisely, the temporal profile recorded in presence (red) or in absence (in blue) of the other channels have been reported. These results show, that due to a careful choice of unmatched group-velocity central wavelengths and a high quality of multiplexing/demultiplexing operation (>20 dB), the overall cross-talks between the different modes appear negligible. In fact, the four picosecond 40-GHz pulse generators act in parallel to each other.

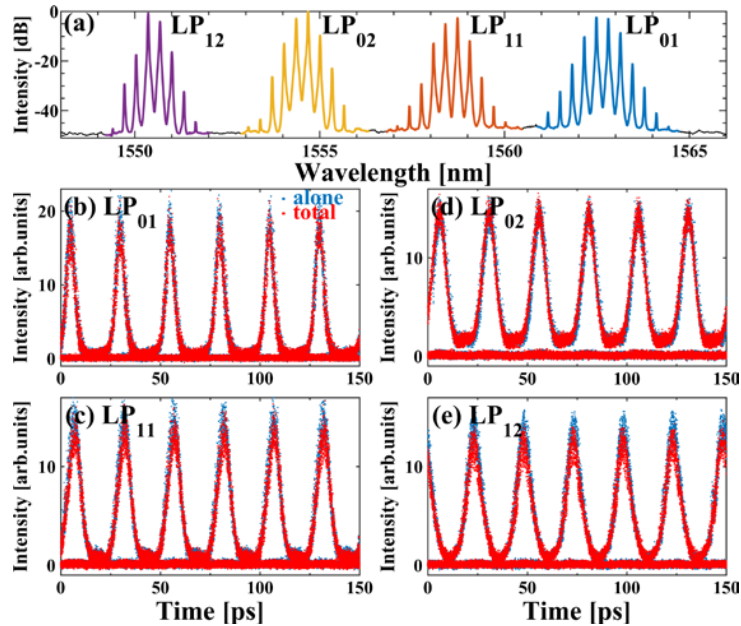

**FIG. 5.** Experimental results. Simultaneous compression of four 40-GHz beat-signals injected respectively into the $LP_{01}$, $LP_{11a}$, $LP_{02}$, and $LP_{12a}$ modes. The output spectra from the four modes are depicted in (a). The demultiplexed intensity profile of the output signals from each mode are shown in (b)-(e), with the blue dots corresponding to single-mode-excitation and the red dots representing the simultaneous excitations of all the four modes.

## 5. Conclusion

In conclusion, we have reported for the first time the generation of four spatially multiplexed picosecond 40-GHz pulse trains in a single 1.8-km long 6-LP multimode fiber. The principle of operation is based on the parallel nonlinear compression of initial beat-signals into well separated pulse trains owing to intra-modal multiple four-wave mixing phenomena. We have successfully generated four spatially multiplexed well-separated 40-GHz picosecond pulse trains in the $LP_{01}$, $LP_{11}$, $LP_{02}$ and $LP_{12}$ of our graded-index few mode fiber with negligible cross-talk. Note that similar results could be obtained in a step-index few mode fiber and could be also extended to a higher number of spatial modes. Moreover, since the group velocity difference between higher order modes is larger within a step-index FMF, it should be even more efficient to remove the deleterious cross-talk effect provided by intermodal cross-phase modulation. Finally, these results demonstrate that spatial multiplexing of nonlinear functionalities can be performed in few-mode fibers and open the path to the development of all-optical processing techniques for spatial division multiplexing applications.

## Funding

This work was founded by the Agence Nationale pour la recherche ANR APOFIS project ANR-17-ERC2-0020-01.

## Acknowledgments

This work benefits from the PICASSO platform in ICB.

## References

1. T. F. Carruthers and I. N. Duling, "10-GHz, 1.3-ps erbium fiber laser employing soliton pulse shortening," Opt. Lett. **21**, 1927-1929, 1996.
2. L. Xu, L. F. K. Lui, P. K. A. Wai, H. Y. Tam, and C. Lu, "40 GHz Actively Mode-Locked Erbium-Doped Fiber Ring Laser Using an Electro-Absorption Modulator and a Linear Optical Amplifier," in OFC 2007, OSA Technical Digest, paper JThA10.
3. A. D. Ellis, R. J. Manning, I. D. Phillips, D. Nesset, "1.6 ps pulse generation at 40 GHz in phase locked ring laser incorporating highly nonlinear fibre for application to 160Gbit/s OTDM networks", Electron. Lett. **35**, 645-646, 1999.
4. B. Bakhshi, and P. A. Andrekson, "40 GHz actively mode locked polarizarion-maintaining erbium fiber ring laser," Electron. Lett. **30**, 411-413, 2000.
5. K. Sato, A. Hirano, and H. Ishii, "Chirp-Compensated 40-GHz Mode-Locked Lasers Integrated with Electroabsorption Modulators and Chirped Gratings," IEEE J. Sel. Top. Quantum Electron. **5**, 590-595, 1999.
6. T. Kobayashi, H. Yao, K. Amano, Y. Fukushima, A. Morimoto, and T. Sueta, "Optical pulse compression using high-frequency electro-optic phase modulation," IEEE J. Quantum Electron. **24**, 382-387, 1988.
7. T. Komukai, Y. Yamamoto, and S. Kawanishi, "Optical pulse generator using phase modulator and linearly chirped fiber Bragg gratings," IEEE Photon. Technol. Lett. **17**, 1746-1748, 2005.
8. V. Torres-Company, J. Lancis, and P. Andres, "Unified approach to describe optical pulse generation by propagation of periodically phase-modulated CW laser light," Opt. Express **14**, 3171-3180, 2006.
9. N. K. Berger, B. Levit, and B. Fischer, "Reshaping periodic light pulses using cascaded uniform fiber Bragg gratings," J. Lightw. Technol. **24**, 2746-2751, 2006.
10. R. Slavik, F. Parmigiani, L. Gruner-Nielsen, D. Jakobsen, S. Herstrom, P. Petropoulos, and D. J. Richardson, "Stable and Efficient Generation of High Repetition Rate (>160 GHz) Subpicosecond Optical Pulses," IEEE Photon. Technol. Lett. **23**, 540-542, 2011.
11. T. Inoue and S. Namiki, "Pulse compression techniques using highly nonlinear fibers," Laser Photonics Rev. **2**, 83-99, 2008.
12. S. V. Chernikov, E. M. Dianov, D. J. Richardson, R. I. Laming, and D. N. Payne, "114 Gbit/s soliton train generation through Raman self-scattering of a dual frequency beat signal in dispersion decreasing optical fiber," Appl. Phy. Lett. **63**, 293-295, 1993.
13. Y. Ozeki, S. Takasaka, T. Inoue, K. Igarashi, J. Hiroishi, R. Sugizaki, M. Sakano and S. Namiki, "Nearly exact optical beat-to-soliton train conversion based on comb-like profiled fiber emulating a polynomial dispersion decreasing profile," IEEE Photonics Technol. Lett. **17**, 1698-1700, 2005.

14. S. V. Chernikov, J. R. Taylor and R. Kashyap, "Experimental demonstration of step-like dispersion profiling in optical fibre for soliton pulse generation and compression," Electron. Lett. **30**, 433-435, 1994.
15. E. A. Swanson, and S. R. Chinn, "40-GHz pulse train generation using soliton compression of a Mach-Zehnder modulator output," IEEE Photon. Technol. Lett. **7**, 114-116, 1995.
16. E. Ciaramella, G. Contestabile, A, D'Errico, C. Loiacono, and M. Presi, "High-power widely tunable 40-GHz pulse source for 160-Gb/s OTDM systems based on nonlinear fiber effects," IEEE Photon. Technol Lett. **16**, 753-755, 2004.
17. S. Trillo, S. Wabnitz, and T. A. B. Kennedy, "Nonlinear dynamics of dual-frequency-pumped multiwave mixing in optical fibers," Phys. Rev. A. **50**, 1732-1747, 1994.
18. J. Fatome, S. Pitois, and G. Millot, "20-GHz to 1-THz repetition rate pulse sources based on multiple four wave mixing in optical fibers", IEEE J. Quantum Electron., **42**, 1038-1046, 2006.
19. J. Fatome, S. Pitois, C. Fortier, B. Kibler, C. Finot, G. Millot, C. Courde, M. Lintz, and E. Samain, "Multiple four-wave mixing in optical fibers: 1.5–3.4-THz femtosecond pulse sources and real-time monitoring of a 20-GHz picosecond source," Opt. Comm. **83**, 2425-2429, 2010.
20. J. J. Pigeon, S. Ya. Tochitsky, and C. Joshi, "High-power, mid-infrared, picosecond pulses generated by compression of a $CO_2$ laser beat-wave in GaAs," Opt. Lett. **40**, 5730-5733, 2015.
21. K. E. Webb, M. Erkintalo, Y. Q. Xu, G. Genty, and S. G. Murdoch, "Efficiency of dispersive wave generation from a dual-frequency beat signal," Opt. Lett. **39**, 5850-5853, 2014.
22. R. J. Essiambre, M. A. Mestre, R. Ryf, A. H. Gnauck, R. W. Tkach, A. R. Chraplyvy, Y. Sun, X. Jiang, and R. Lingle, Jr. "Experimental observation of inter-modal cross-phase modulation in few-mode fibers," IEEE Photonics Technol. Lett. **25**, 535-538, 2013.
23. R. J. Essiambre, M. A. Mestre, R. Ryf, A. H. Gnauck, R. W. Tkach, A. R. Chraplyvy, Y. Sun, X. Jiang, and R. Lingle, "Experimental investigation of inter-modal four-wave mixing in few-mode fibers," IEEE Photon. Technol. Lett. **25**, 539-542, 2013.
24. S. M. M. Friis, I. Begleris, Y. Jung, K. Rottwitt, P. Petropoulos, D. J. Richardson, P. Horak, and F. Parmigiani, "Inter-modal four-wave mixing study in a two-mode fiber," Opt. Express **24**, 30338-30349, 2016.
25. H. Zhang, M. Bigot-Astruc, L. Bigot, P. Sillard, and J. Fatome, "Multiple modal and wavelength conversion process of a 10-Gbit/s signal in a 6-LP-mode fiber," Opt. Express **27**, 15413-15425, 2019.
26. G. Labroille, P. Jian, N. Barré, B. Denolle, and J. Morizur, "Mode Selective 10-Mode Multiplexer based on Multi-Plane Light Conversion," in OFC 2016, paper Th3E5.
27. P. Sillard, D. Molin, M. Bigot-Astruc, K. de Jongh, and F. Achten, "Micro-Bend-Resistant Low-DMGD 6-LP-Mode Fiber," in OFC 2016, paper Th1J5.
28. R. Dupiol, A. Bendahmane, K. Krupa, J. Fatome, A. Tonello, M. Fabert, V. Couderc, S. Wabnitz, and G. Millot, "Intermodal modulational instability in graded-index multimode optical fibers," Opt. Lett. **17**, 3419-3422, 2017.